\documentclass[aps,twocolumn,superscriptaddress,showpacs,]{revtex4}
\usepackage{graphicx}

\begin{document}

\newcommand{\ud}{\,d}

\newcommand{\be}{\begin{equation}}
\newcommand{\ee}{\end{equation}}
\newcommand{\bea}{\begin{eqnarray}}
\newcommand{\eea}{\end{eqnarray}}


\title{Hybrid Quintessential Inflation}


\author{Mar~Bastero-Gil}
\affiliation{%
Departamento de F\'{\i}sica Te\'orica y del Cosmos,
Universidad de Granada, Granada-18071, Spain}%

\author{Arjun~Berera}
\affiliation{
School of Physics and Astronomy, University of Edinburgh, Edinburgh EH9 3JZ, UK}

\author{Brendan M. Jackson}
\affiliation{Institute for Astronomy, School of Physics and Astronomy,
University of Edinburgh, Edinburgh EH9 3HJ, UK}

\author{Andy Taylor}
\affiliation{Institute for Astronomy, School of Physics and Astronomy,
University of Edinburgh, Edinburgh EH9 3HJ, UK}

\affiliation{}


\date{\today}

\begin{abstract}
A model is presented in which a single scalar field is responsible
for both primordial inflation at early times and then dark energy
at late times.  This field is coupled to a second scalar field which
becomes unstable and starts to oscillate after primordial inflation,
thus driving a reheating phase that can create a high post-inflation
temperature.  This model easily avoids overproduction of gravity waves,
which is a problem in the original quintessential inflation model
in which reheating occurs via gravitational particle
production. 
\end{abstract}



\maketitle



\section{Introduction}

Peebles and Vilenkin proposed in \cite{Peebles:1998qn} that both
inflation and dark energy could be a result of the same scalar field,
with vacuum expectation value $\phi$, interacting only with gravity
and itself via the potential term $V(\phi)$, which they chose to be 
\begin{eqnarray}
V\left(\phi \right) = & \lambda \left(\phi^4 + M^4\right) & \quad \text{for} \quad \phi <0 \nonumber \\
  = & \displaystyle{\frac{\lambda M^8}{\phi^4 + M^4}}  & \quad \text{for} \quad \phi \ge 0.
\label{phi_potential}
\end{eqnarray} 
At tree level the evolution of the vacuum
expectation value of the $\phi$-field is governed by
\begin{equation}
\ddot{\phi} + 3\frac{\dot{a}}{a} \dot{\phi} = -\frac{\partial
  V}{\partial \phi } \,,  
\end{equation}
with the cosmological expansion rate related to energy density $\rho$ by
the Friedmann equation, 
\begin{equation} \label{basicFriedmann}
H^2 = \left(\frac{\dot{a}}{a}\right)^2 
 = \frac{8 \pi}{3 m_{\rm pl}^2} \rho \,,
\end{equation}
where the Planck mass $m_{\rm pl} = G^{-1/2} = 1.22 \times 10^{19} $ GeV 
and overdots signify derivatives with respect to coordinate time.

In the original scenario \cite{Peebles:1998qn}, the universe begins
dominated by the potential energy of the scalar field $\phi$. The field has
some large, negative value, and slowly rolls toward the origin. The
slow-roll conditions of inflation are satisfied, and the universe
expands exponentially.  During inflation, fluctuations in $\phi$ are
frozen into the field. These seed future structure in the universe. To
provide the correct level of fluctuations, as usual $\lambda \simeq
10^{-14}$.  When $\phi \sim-m_{\rm pl}$, the inflationary epoch draws to a
close. Thus far the situation is identical to $\phi^4$ chaotic
inflation \cite{chaotic}, but from here on it differs.  The kinetic
energy $\frac{1}{2} \dot{\phi}^2$ of the field is no longer
negligible, and it soon begins to dominate the universe
\cite{Spokoiny:1993kt}. This phase is termed ``kination''. The field
behaves approximately as stiff matter, with its energy density
redshifting as $\rho_\phi \propto a^{-6}$. Taking the potential energy
at the end of inflation to be the same as the kinetic energy, a simple
estimate gives
\begin{equation}\label{Friedmann}
H^2 \simeq \frac{8\pi}{3} \lambda m_{\rm pl}^2 \left( a_x/a \right)^6 \,,
\end{equation}
with solution, $a \propto t^{1/3}$ and thus $t = \left(3H
\right)^{-1}$.  Here the subscript $x$ indicates the value at the end
of inflation. During this kinetic dominated phase, the field moves as
\begin{equation} \label{phi_equation}
\phi = \sqrt{\frac{6}{8\pi}} m_{\rm pl}\ln\left(a/a_x\right) - m_{\rm pl} \,.
\end{equation}
Thus, spacetime started in approximately de Sitter form, and ended dominated
by the kinetic energy of a homogeneous field. Ford showed \cite{Ford:1986sy}
this transition leads to gravitational particle production. 
The result is a small but important energy density of relativistic
particles, 
\be
\rho_{\rm r} \simeq 0.01 N_{\rm s}H_x^4 \,,
\ee
where $N_{\rm{s}}$ is the number of scalar fields.
Thermalization was found in \cite{Peebles:1998qn} to occur at
a radiation temperature 
\begin{equation}
T_{\rm{th}} \simeq 10^9 N_{\rm{s}}^{3/4} \; \rm{GeV} \,.
\end{equation}
The radiation redshifts away slower than the energy in the
field. Provided $M$ in Eq. (\ref{phi_potential})
is not too large, Peebles and Vilenkin showed the
universe transitions to a radiation dominated epoch, with a
temperature of
\begin{equation}
T_{\rm{RH}} \simeq 10^3 N_{\rm{s}}^{3/4} \; \rm{GeV} \,.
\end{equation}
The field remains essentially static from this point on, mimicking a
cosmological constant. The value of $M$ can be then chosen so that 
$V(\phi_{\rm{r}})$ matches today's observed value of dark energy density.
In the analysis in \cite{Peebles:1998qn}, this turned out to be $M \sim
10^6$ GeV. 

Unlike most inflationary scenarios, in the one by Peebles and Vilenkin
\cite{Peebles:1998qn} at the end of inflation
the inflaton field does not undergo a series of
damped oscillations about a potential minima, which is the mechanism
by which reheating usually takes place \cite{Albrecht:1982mp}.  The
absence of this behaviour means gravitational particle production has
to be relied upon instead. The same gravitational mechanism works to
produce a stochastic background of gravitational waves (GW)
\cite{Starobinsky:1979ty,Allen:1987bk,Sahni:1990tx}, and the
overproduction of gravity waves is one of the potential dangers of
this model. Gravitons behave as minimally massless scalar fields, and
the energy density of the gravitons by the end of inflation is just
that of a single scalar field, times two polarization states, so that
the ratio of the energy densities in GW to radiation at the end of
inflation is simply $(\rho_{GW}/\rho_{\rm r})_x \simeq 2/N_{\rm s}$. At the
time of nucleosynthesis $\rho_{GW}$ will contribute as an effective
extra degree of freedom in radiation, but the success of Big Bang
Nucleosynthesis (BBN) in predicting the abundances of light elements
puts a constraint on the extra number of light degrees of freedom, be
those neutrinos species or gravitons, such that
$(\rho_{GW}/\rho_{\rm r})_{BBN} < 0.2$; tracing back the evolution of this
ratio upto the start of the kination period this would sets a lower
bound $N_{\rm s} \gtrsim 100$ \cite{Peebles:1998qn}.

The problem of gravity waves overproduction is quite generic
in models of inflation followed by a long period of kination
\cite{Sahni:2001qp}, or in general stiff matter domination
\cite{Giovannini:1998bp},   
like brane world inflation \cite{braneworld}. To allow for a more
effective reheating process, able to suppress the relative contribution
of the GW at the time of BBN, one can invoke alternative methods like
instant preheating \cite{instantpre}, curvaton reheating
\cite{curvreh}, or Born-Infeld reheating \cite{BIreh}.  Any
alternative implies introducing extra scalar degrees of freedom at the
time of inflation, like in curvaton reheating, and/or direct couplings of the
inflaton field to the light degrees of freedom as in instant
preheating. 

In this letter we explore the possibility of a simpler
scenario, recovering a more typical reheating mechanism driven by the
decay of an oscillating massive field \cite{Albrecht:1982mp}. We
extend Peebles and Vilenkin model by a new scalar field $\chi$ coupled
to the inflaton field \cite{abwhepp} with a hybrid-like potential 
\cite{Linde:1993cn}. In our scenario, once the inflaton
field falls below a critical 
value, the $\chi$ field can start oscillating, thus gaining energy
that afterwards can be converted into radiation through perturbative
decay, as in the usual reheating mechanism. However, in
contrast to the standard reheating picture, in our scenario
reheating takes place during kination instead of the more standard
matter domination. Unless the perturbative decay of the $\chi$ field
is tiny, this results in a larger reheating $T$ than 
in the original model of Peebles and Vilenkin \cite{Peebles:1998qn},
and a shorter kination phase. The more efficient reheating also
ensures that radiation domination takes over kination,
well before the inflaton vacuum energy starts dominating again.  

This letter is organized as follows.  In Section II the
potential and parameters of the model are set.
Also the general
behaviour is described of the field $\chi$ after inflation, 
when it can oscillate and drive reheating.
The reheating temperature $T_{\rm RH}$ is then
computed in Section III. For the mechanism to work, we need to check first
that $\chi$ indeed oscillates when the inflaton field passes through
the critical point, and that it does not  backreact on the evolution
of $\phi$. Fulfilling these conditions sets the constraints on the
model parameters, which are given in Section IV. In Section V we
present the range of $T_{\rm RH}$ consistent with the constraints. The
transition to a radiation dominated universe is studied in Section
VI. Once the constraints are fulfilled, the transition before the
onset of dark energy domination is practically ensured. As
this is a 
hybrid-like model, we comment on the issue of domain
walls in section VII. Finally in Section VIII we present the summary and future
work related to quantum corrections.     

\section{General behaviour of the hybrid field}

To enable once more the standard reheating, we
introduce a new scalar field $\chi$, that we couple to the inflaton
field. The effective potential at tree-level is
\begin{equation}
\label{U_potential}
U \left(\phi,\chi\right) = V(\phi) + \frac{g^2}{2} \chi^2 \left(\phi^2
- m^2 \right) + \frac{\lambda_\chi}{4} \chi^4 \,,
\end{equation}
where $V( \phi )$ is the potential given by equation
(\ref{phi_potential}). The parameters $g$, $m$, and $\lambda_\chi$ are
as yet undetermined constants. Note that when $\chi$ is relaxed near
the origin, $U(\phi,\chi) \approx V(\phi)$.

The $\chi$-field is assumed to be located somewhere near its minima at 
$\chi = 0$. As the $\phi$ field evolves from large negative values to large
positive ones, the turning point at the origin temporarily becomes
unstable for the $\chi$-field, and two minima are generated 
on either side. The position of
the minima, $\chi_{\rm{min}}$ are given by
\begin{equation}
\label{minima_location}
\chi_{\rm{min}}^2 = \frac{g^2}{\lambda_{\chi}} \left(m^2 - \phi^2\right) \,.
\end{equation}
They exist only while $\left|\phi\right| < m$.  The bottom of the
well is at a negative value of potential energy. This is a result of
how the energy of the system has been defined.  It has no physical
consequence for the present scenario,
since the total energy density will always remain positive,
as it is
dominated by the kinetic inflaton energy density. 

The $\chi$-field has a characteristic response time $\tau$ to react to
changes in the potential. This can be estimated as $\tau^{-1} \sim
m_{\chi}(\phi) \sim g(m^2 - \phi^2)^{1/2} $, provided $\phi$ is not
too close to $m$. If changes in the potential take place quicker than
this response time, the field will have no chance to react. For
instance, if $\phi$ moves between $-m$ and $+m$ in a time $\Delta t
\ll \tau$ throughout, $\chi$ will have no time to move before
the origin becomes a stable minima once more. On the other hand, if
the changes take place on timescales that are much longer than the
response time, i.e. $\Delta t \gg \tau$, the $\chi$ field
will be able to relax into the minima very quickly. If this is the
case throughout, there will hardly be any oscillations (and hardly any
reheating).

A fact alleviates the above difficulties: $\phi$ is slowing down, and
is doing so quickly (its evolution is logarithmic in time). This means
the timescale $\Delta t$ becomes gradually longer. Furthermore, the
timescale $\tau$ that governs how quickly $\chi$ reacts is not
static. It is longest when $\left| \phi \right | \approx m $ and is
shortest at $\phi = 0$. We will take the view that the field begins to
move before $\phi \approx 0$. This places an immediate constraint on
the model parameters: $\Delta t \gg \tau|_{\phi = 0}$. We should
immediately note the better this inequality is satisfied, the sooner
$\chi$ will start to move after becoming unstable. This will limit the
amount of energy for its oscillations.

As a simplified picture, consider the $\chi$ field to be
frozen at the origin up until a time $\tau$ after becoming unstable.
Afterwards, the potential will be approximately constant during
the fast oscillations of the field. The energy in the oscillations
(available for reheating) will then be the depth of the potential well
at the point when the field begins to move.  
This will be some fraction $f(g,m)$ of
the maximum potential well depth. The energy available for reheating
can then be written:
\begin{equation} \label{initial energy}
\rho_{\chi}^{(0)} \simeq f\left(g,m\right) \frac{m^4 g^4}{4\lambda_{\chi}} \,.
\end{equation}
Eventually, after many oscillations, $\phi$ reaches $+m$, and the
$\chi$ - field then oscillates about its stable minima at the origin.
These oscillations will reheat the universe.



The fraction $f(g,m)$ now needs to be estimated.  For this, we
assume the $\chi$ -field moves some short 
time after it becomes unstable, with the instability occurring
at $\phi = -m$. As this time signals the start of
reheating, it  will be called $t_{\rm{re}}$.  Writing the time since
the start of reheating as $\delta t$, then we can estimate that the
$\chi$ field will begin to move when $\tau \sim \delta t$. 
In other words, the
field moves after it has been unstable for a time approximately the
same as its characteristic reaction time (which is itself a function
of $\delta t$). 
The depth of the well is given by
\begin{equation}
U_{\rm{min}} = m^{-4}\left( m^2 - \phi^2 \right)^2 \frac{m^4
g^4}{4 \lambda_{\chi}} \,.
\end{equation}
Comparison with equation (\ref{initial energy}) shows 
\be
f(g,m) = m^{-4}\left(m^2 - \phi^2 \right)^2 \,,
\ee
with $\phi$ evaluated when $\delta t \simeq \tau$.
We assume $\phi$ moves only slightly, an amount $\delta \phi =\phi -m$, before
$\chi$ begins to move.  Re-writing equation (\ref{phi_equation}) gives
\begin{equation} \label{phit}
\delta \phi \approx \frac{1}{3}\sqrt{\frac{6}{8 \pi}} m_{\rm{pl}} \frac{\delta
  t}{t_{\rm{re}}} \,.
\end{equation}
We wish to find the value of $\delta t$ satisfying $\delta t \simeq \tau$. As $\tau^{-1} \simeq  g(m^2 - \phi^2)^{1/2}$, we can insert $\phi
= \delta \phi - m$ 
to find 
\begin{equation}
\tau^{-2} \sim m g^2 \frac{2}{3} \sqrt{\frac{6}{8 \pi}} m_{\rm{pl}}
\frac{\delta t}{t_{\rm{re}}} \,.
\end{equation}
The same substitution into $f(g,m)$ yields
\begin{equation}
f(g,m) \simeq  4 \frac{(\delta \phi)^2}{m^2} \,,
\end{equation}
to leading order in $\delta \phi$.

Applying the condition $\delta t \simeq \tau$, we can find the
value of $\delta t$ when the $\chi$ - field
begins to move. Relating $\delta \phi$ to $\delta t$ with equation 
(\ref{phit}), we finally obtain 
\begin{equation} \label{fraction}
f(g,m) \simeq  \left(\frac{1}{3 \pi}\right)^{2/3}
             \left( \frac{m_{\rm{pl}}}{m} \right)^{8/3} 
             \left(m_{\rm{pl}}t_{\rm{re}}\right)^{-4/3}
             g^{-4/3}\,.
\label{fgm}
\end{equation}
Note that this order of magnitude approximation breaks down if $\chi$
begins moving when $\delta \phi$ is not small compared to $m$,
signalled by $f(g,m)$ approaching (or exceeding) unity. This
expression should not be trusted in such circumstances. Assuming
$\delta \phi$ to be small  is a slightly  stronger constraint than the
constraint discussed earlier, that 
$\Delta t \gg \tau$. The former gives the condition the field moves
\emph{quickly} after becoming unstable, the latter is just the
condition the field moves \emph{at all}, at or before $\phi \approx 0$. We
will  return to this point when discussing the parameter constraints in
Section IV.  

\section{Reheating temperature}

Once $\phi > m$, the $\chi$ field will return to the origin. If
it acquired kinetic energy due to its temporary displacement, it will now
oscillate about the origin. The previous section established the
amount of energy expected from these oscillations. 

If the oscillations are small, the field can be approximated as
undergoing simple harmonic motion. In such a case, the elementary
theory of reheating can be applied \cite{Albrecht:1982mp}. 
In this phenomenological approach,
an extra term $\Gamma_{\chi}\dot{\chi}$ is added to the equation of
motion of the field to account for particle decay. The value of
$\Gamma_{\chi}$ is taken to be the decay rate of the particle. The
field $\chi$ then obeys the equation of motion
\begin{equation} \label{chi_motion}
 \ddot{\chi} + 3\frac{\dot{a}}{a} \dot{\chi} + \Gamma_{\chi}\dot{\chi}
       = -\frac{\partial U}{\partial\chi} \,.
\end{equation}
This approach is valid when the oscillations are small and the
oscillations are well approximated as simple harmonic motion. For a
more complete picture, valid at the early stages we should also
consider the effect of preheating \cite{preheating} (see also
\cite{instantpre} for an analysis of preheating in the context of
quintessential inflation). However,
these details will be ignored in this paper
and we will examine only the simplest
reheating estimates.

We stress that this phenomenological approach is only valid while
the field is undergoing coherent oscillations about its minima. It
will not be valid otherwise, nor over short timescales, and is likely
to fail when $\left|\phi \right| \sim m$. For this reason we will
consider reheating to take place only while $\phi > m$, so that we may
have confidence our calculations are always carried out in an appropriate
regime. 

We assume $H$, $\phi$ and $\dot \phi$ can be taken to be approximately
constant over a single oscillation. Equation (\ref{chi_motion}) is
re-written by replacing $\dot{\chi}^2$ by its value over a complete
oscillation, $\langle \dot{\chi}^2 \rangle_{cycle} = \rho_{\chi}$,
which is valid for
simple harmonic motion.  This yields
\begin{equation}
\label{chi_decay}
\dot{\rho_{\chi}} + 3H \rho_{\chi} + \Gamma_{\chi}\rho_{\chi} = g^2
\langle \chi^2 \rangle \phi \dot{\phi} \,.
\end{equation}
Assuming the field is still undergoing simple harmonic motion,
$\frac{1}{2} \rho_{\chi} = V(\phi,\chi) = \frac{1}{2} g^2 \phi^2
\langle \chi^2 \rangle$. Then we can write
\begin{equation}
\label{chi_late_time}
\dot{\rho_{\chi}} + 3H \rho_{\chi} + \Gamma_{\chi}\rho_{\chi} =
\frac{\dot{\phi}}{\phi} \rho_{\chi} \,.
\end{equation}
This can be solved,
\begin{equation} \label{chi_energy}
\rho_{\chi} = \rho_{\chi}^{(m)} \left( \frac{a_{{m}}}{a} \right)^{3}  
              \frac{\phi(t)}{m}\exp{\left[-\Gamma_{\chi}(t-t_{m})\right]} \,, 
\end{equation}
with
 \begin{equation} \label{phim}
 \frac{\phi}{m} = \sqrt{\frac{6}{8 \pi}}
\frac{m_{\rm{pl}}}{m} \ln\left(\frac{a}{a_{{m}}}\right) + 1 \,.
 \end{equation}
We have used subscript ${m}$ to indicate the value of a variable when
$\phi = m$.
From energy conservation, it follows that the radiation density must obey
\begin{equation}\label{radiation_equation}
\dot{\rho}_{r} + 4H \rho_{\rm r} = \Gamma_{\chi}\rho_{\chi} \,.
\end{equation}
Imposing the condition there is no radiation at the start of decay, an
approximate solution is found by neglecting the exponential decay of
$\rho_{\chi}$. This will be valid up to $t \approx
\Gamma_{\chi}^{-1}$. After this, the energy in the $\chi$ field will
decay rapidly away and the radiation will simply redshift with its
usual $a^{-4}$ behaviour.

Inserting our earlier expression for $\rho_{\chi}$ into 
equation (\ref{radiation_equation}), the solution
for the radiation energy density can be written as
\begin{eqnarray} \label{radiation_solution2}
\rho_{\rm r} = \frac{3}{4} \rho_\chi^{(m)} \Gamma_{\chi} t_{{m}}\Big[
&\big(1 -  \left( t/t_{{m}} \right)^{-4/3} 
\big ) \big ( 1-b \big) \nonumber \\
& +\, b \left(4/3\right) \ln(t/t_{{m}}) \, \Big],  
\end{eqnarray}
with $b = \frac{1}{4} \sqrt{\frac{6}{8\pi}}
\frac{m_{\rm{pl}}}{m}$ and the constant of integration chosen so that
there is no radiation at $\phi = m$.
 
Assuming $(t/t_{\rm re})^{-4/3} \ll 1$ before $t$ reaches
$\Gamma_{\chi}^{-1}$, the energy density in radiation should
be well approximated by 
\begin{equation}
\label{rhorsol}
\rho_{\rm r} \simeq \frac{3}{4} \rho_\chi^{(m)} \Gamma_{\chi} t_{{m}} \left[
1 - b + \frac{4}{3} b \ln\left(t/t_{{m}} \right) \right ] \,.
\end{equation}
The energy density of the radiation continues growing logarithmically,
despite the energy loss from redshifting. This is due to the mild
amount of energy being added to the $\chi$ field by its
coupling to $\phi$. It will continue to grow in this way until 
$t \approx \Gamma_{\chi}^{-1}$. If this is a sufficiently late time that the
last term in Eq. (\ref{rhorsol}) dominates, then
\begin{equation}
\rho_{\rm r} \sim  \frac{1}{4} \sqrt{\frac{6}{8\pi}}\frac{m_{\rm{pl}}}{m} 
\rho_\chi^{(m)} \Gamma_{\chi} t_{{m}} \ln(t/t_{{m}}) \,.
\label{rhor}
\end{equation}
Once the universe has had chance to thermalize, the temperature is
related to the energy density via,
\begin{equation}
\rho_{\rm r} = \frac{g_* \pi^2 T^4}{30} \,,
\end{equation}
where $g_*$ is the number of degrees of freedom. 

\begin{figure}
\includegraphics[width=0.5\textwidth]{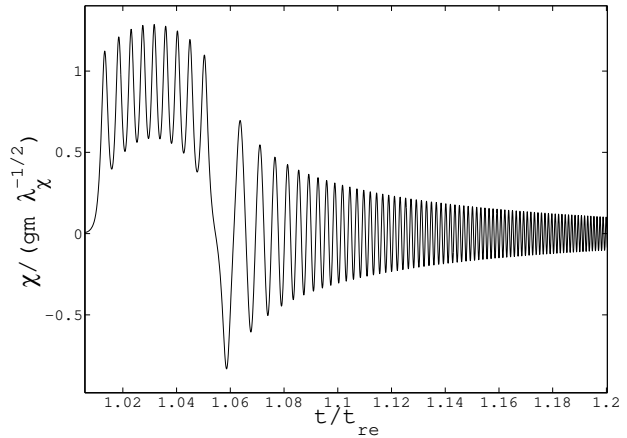}
\caption{\label{figure2} Numerical solution for the motion of the
$\chi$ field, with $g = 5 \times 10^{-4}$, $\lambda_{\chi} = 1$, $m =
5 \times 10^{16}$ GeV and $\alpha = 1 \times 10^{-4}$. The field
begins at rest, with a small displacement from the origin. The
resulting oscillations continue when the field returns to oscillate
about the origin, and are well approximated as damped, simple harmonic
motion.}
\end{figure}

Thermalization occurs when the interaction rate $n_r \sigma$ becomes
comparable to the expansion rate $H$, where $n_r$ is the number
density of the light degrees of freedom and $\sigma$ the their
interaction cross section. We can estimate that the light degrees of
freedom are created with a typical energy $\omega \sim
\rho_{\rm r}^{1/4}(a_{\rm{re}}/a)$, and $\sigma \sim \alpha_g/\omega^2$,
with $\alpha_g$ being the strength of the mediating interactions.
Using Eq. (\ref{rhor}) with the expression of $t_{\rm{re}} \simeq
t_{m}$ given in the next section Eq. (\ref{reheating_time}), and
as a typical value for the coupling in the
cross-section $ \alpha_g \sim 0.01$, one gets that at the beginning of
the reheating period, 
\be 
\frac{n_r \sigma}{H_m} \sim \frac{\rho_{\rm r}^{1/4} \alpha_g}{H_m} \sim 
10^5 (10 \alpha^{1/4})\left(\frac{m}{m_{\rm{pl}}}\right)^{1/3}
\left(\frac{g^{11/3}}{\lambda_\chi}\right)^{1/4}\,.  
\label{thermal}
\ee 
We have also written the  decay rate for a massive particle as
$\Gamma_\chi \simeq \alpha m_\chi$, where $\alpha$ is the coupling
constant mediating the decay, and $m_\chi \simeq g m$ the $\chi$ mass. 
For the analyses of the reheating done in this section, the decay rate
and therefore the coupling $\alpha$ must be such that $\Gamma_\chi
t_{m} \ll 1$. For values of the parameters consistent with the
constraints given in the following section, this is general the case
with $\alpha \simeq 10^{-4}$. Significantly, with this choice for
$\alpha$ then Eq. (\ref{thermal}) is
already larger than one. Therefore, light degrees of freedom
thermalized promptly after they are produced.

We have confirmed these approximations are successful in their
appropriate regimes by numerically solving this system of differential
equations (Friedmann's equation, the equations of motion for both
fields, and the radiation energy density).  Figure \ref{figure2} shows
the numerically determined evolution of the $\chi$ - field as it
becomes unstable, and the subsequent damped oscillations that reheat
the universe.

\section{Parameter constraints}

We have made two assumptions that can be formulated as simple
constraints on combinations of parameters. The first of these is that
$\chi$ begins to move toward its new equilibrium at or before $\phi
\approx 0$. Earlier we noted this condition was $\Delta t \gg \tau$. The second
assumption is that the $\chi$ field does not significantly influence
the motion of the $\phi$ field. It could do this either from the
coupling, or from the energy density of the field modifying the
expansion rate of the universe.

First we shall calculate $\Delta t$. As $\phi = -m$ at $t_{\rm{re}}$, we
can use equation (\ref{phim}) with these values inserted. Re-arranging,
\begin{equation}
t_{m} = t_{\rm{re}} e^{\sqrt{48 \pi} \frac{m}{m_{\rm pl}}}\,,
\end{equation}
and by writing $\Delta t = t_{{m}} - t_{\rm{re}}$, gives
\begin{equation}
\Delta t  = t_{\rm{re}} \left(e^{\sqrt{48\pi}\frac{m}{m_{\rm pl}}} -1 \right)
\simeq t_{\rm{re}} \sqrt{48\pi}{\frac{m}{m_{\rm pl}}} \,,
\end{equation}
with the rightmost expression
in the case $m/m_{\rm pl} \ll 1$. Thus, it is noteworthy that $t_{\rm{re}} \approx
t_{{m}}$ is a good approximation.

An estimate of $t_{\rm{re}}$ is still needed. We use equation
(\ref{Friedmann}) and $H = \frac{1}{3} t^{-1}$, to write it in terms
of a ratio of scale-factors, 
\begin{equation}
\label{reheating_time}
t_{\rm{re}} \simeq \left(\frac{a_{\rm re}}{a_x}\right)^3
\frac{1}{\sqrt{24\pi \lambda m_{\rm pl}^2}} \,, 
\label{tre}
\end{equation}
where $a_{\rm{re}}$ is the scale factor at $t_{\rm{re}}$. 
The time for $\phi$ to move between $-m$ and $+m$ is then given by 
\begin{equation}
\Delta t \simeq \left(\frac{a_{\rm re}}{a_x} \right)^3 \sqrt{\frac{2}{\lambda}}
\frac {m}{m_{\rm pl}^2} \,.
\end{equation}
The ratio of scale-factors can be found from equation (\ref{phi_equation}). 
Re-arranging and setting $\phi = -m$ at $a=a_{\rm re}$,
\begin{equation}
a_{\rm re}/a_x = e^{\left(1-m/m_{\rm pl}\right)\sqrt{\frac{8\pi}{6}}} \simeq 8 \,.
\end{equation}
As such the time interval available can be written simply as
\begin{eqnarray}
\label{time}
\Delta t 
         & \simeq & 10^{-9}\frac{m}{m_{\rm pl}}  \left(\text{GeV}\right)^{-1} \,,
\end{eqnarray}
with $\lambda = 10^{-14}$.  
Recalling $\Delta t \gg \tau$, a constraint on the model 
parameters can be constructed:
\begin{equation}
\left(\frac{m}{m_{\rm pl}} \right)^2 g \gg \; 10^{-10} \,.
\label{cdeltat}
\end{equation}
Now that we have an expression for $t_{\rm{re}}$, a constraint
on $f(g,m) \ll 1$ can also be calculated. Replacing Eq. (\ref{tre}) into
(\ref{fgm}) gives
\begin{equation}
\left(\frac{m}{m_{\rm pl}} \right)^2 g \gg 2
\sqrt{2 \lambda}\left(\frac{a_x}{a_{\rm re}} \right) \simeq 10^{-9}\, . 
\end{equation}
This is comparable although slightly more restrictive than
Eq. (\ref{cdeltat}). 
This suggests that there can be a region in parameter space
where the second constraint is violated (suggesting $1 \gtrsim f(g,m)
\gtrsim 0.1$) but the first satisfied (so that the field still begins
to move out of its unstable position).  The resulting reheating
temperature in this region is also fairly insensitive to changes in $g$
and $m$, compared with when the second constraint is well satisfied
(and the reheat temperature influenced by $f(g,m)$ as given by
equation (\ref{fraction})). Nevertheless, to ensure that we are in the
region of parameter space for which the field has
oscillated enough to reheat the universe, when referring to these
constraints we will use $(m/m_{\rm pl})^2 g \geq  10^{-8}$. Having
$m_{\rm pl}$ as the largest possible mass scale in the model , the
strength of the $\phi$-$\chi$ interaction is therefore bounded from
below with $g \geq 10^{-8}$.

A second constraint exists from the requirement that the $\chi$ field
does not influence the motion of the $\phi$ field. First we will
consider the requirement the negative potential energy from the
coupling term is not significant compared to the kinetic energy of
$\phi$. As the kinetic energy is constantly diminishing, it will be
simpler to overestimate the potential energy and underestimate the
kinetic energy. We will therefore take the potential energy to be its maximum
value, and the kinetic energy to be the value at $\phi = m$. No matter
the evolution of these quantities, if the quantities evaluated at
these two time intervals are not comparable, they never will be. Reading
off the kinetic energy from equation (\ref{Friedmann}) and requiring this
always greatly exceeds the maximum of the negative potential energy
gives,
\begin{equation}
\lambda m_{\rm{pl}}^4 \left( \frac{a_x}{a_m} \right)^6 > \frac{m^4
  g^4}{4 \lambda_{\chi}} \,,
\end{equation} 
or
\begin{equation}
\left( \frac{m}{m_{\rm{pl}}}\right)^4 \frac{g^4}{\lambda_{\chi}} < 10^{-19} \,.
\end{equation}
If we choose values to satisfy $\Delta t \gg \tau$, 
\begin{equation} 
\label{energy_constraint}
\frac{g^2}{\lambda_{\chi}} \ll 10^{-3} \,.
\end{equation}

Now consider the effect of the coupling on the motion of
$\phi$ directly. The equation of motion gives
\begin{equation}
\ddot{\phi} + 3H\dot{\phi} - g^2 \chi^2 \phi = 0 \,,
\end{equation}
where the kinetic energy is treated as the dominant contribution
to energy of the $\phi$ field. The third term is the contribution due
to the coupling, and we wish to ensure this is negligible compared to
the second. Using equation (\ref{basicFriedmann}) to express Hubble's
parameter in terms of the kinetic energy 
$\rho_{\phi} \approx 1/2\dot{\phi}^2$, this condition
can be written as,
\begin{equation}
\sqrt{\frac{4 \pi}{3}} \frac{6}{m_{\rm{pl}}} \rho_{\phi} \gg g^2
\chi^2 \phi \,. 
\end{equation}
\begin{widetext}

\begin{figure}[t]
\begin{tabular}{cc}
\includegraphics[width=0.5\textwidth]{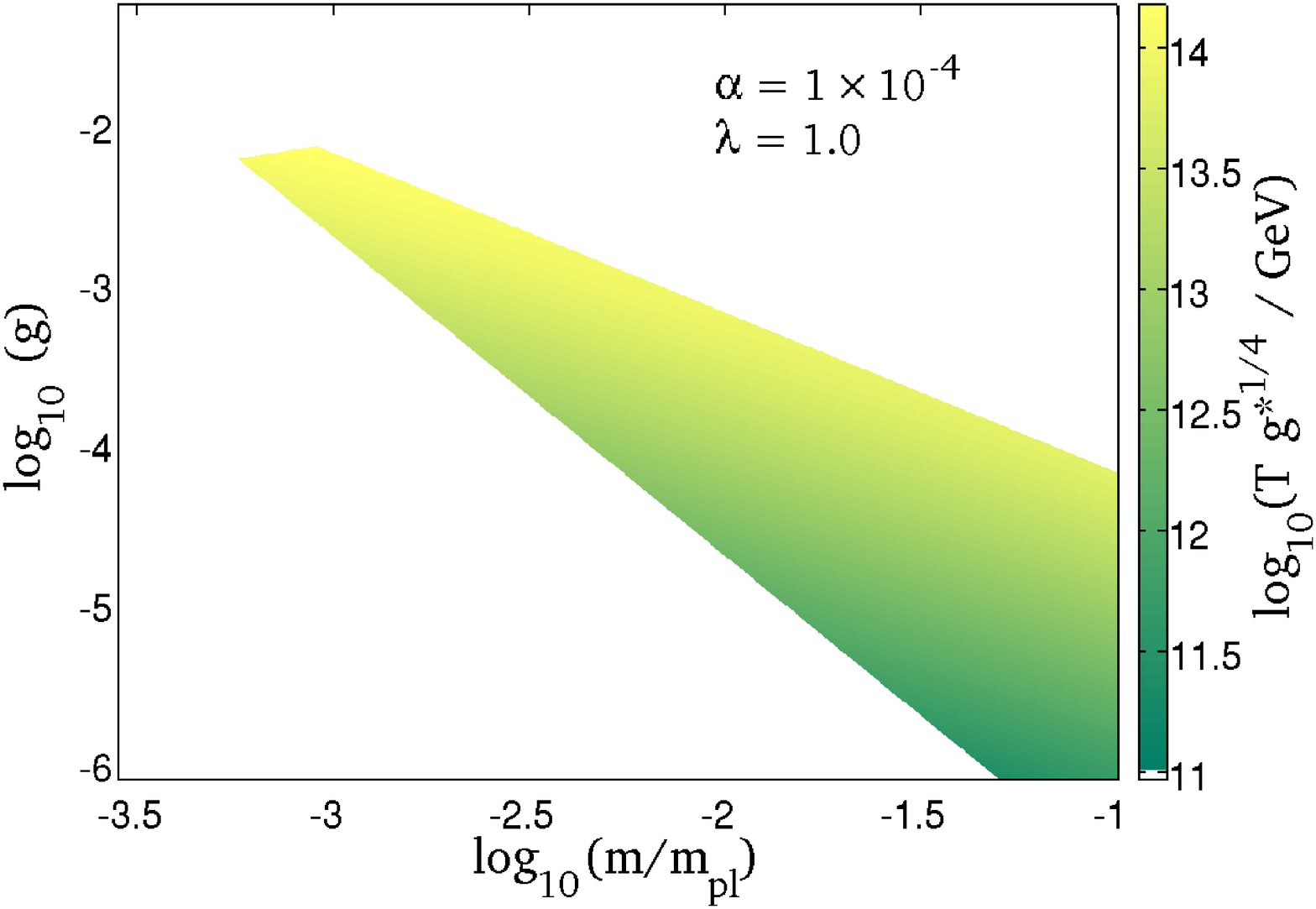} & 
\includegraphics[width=0.5\textwidth]{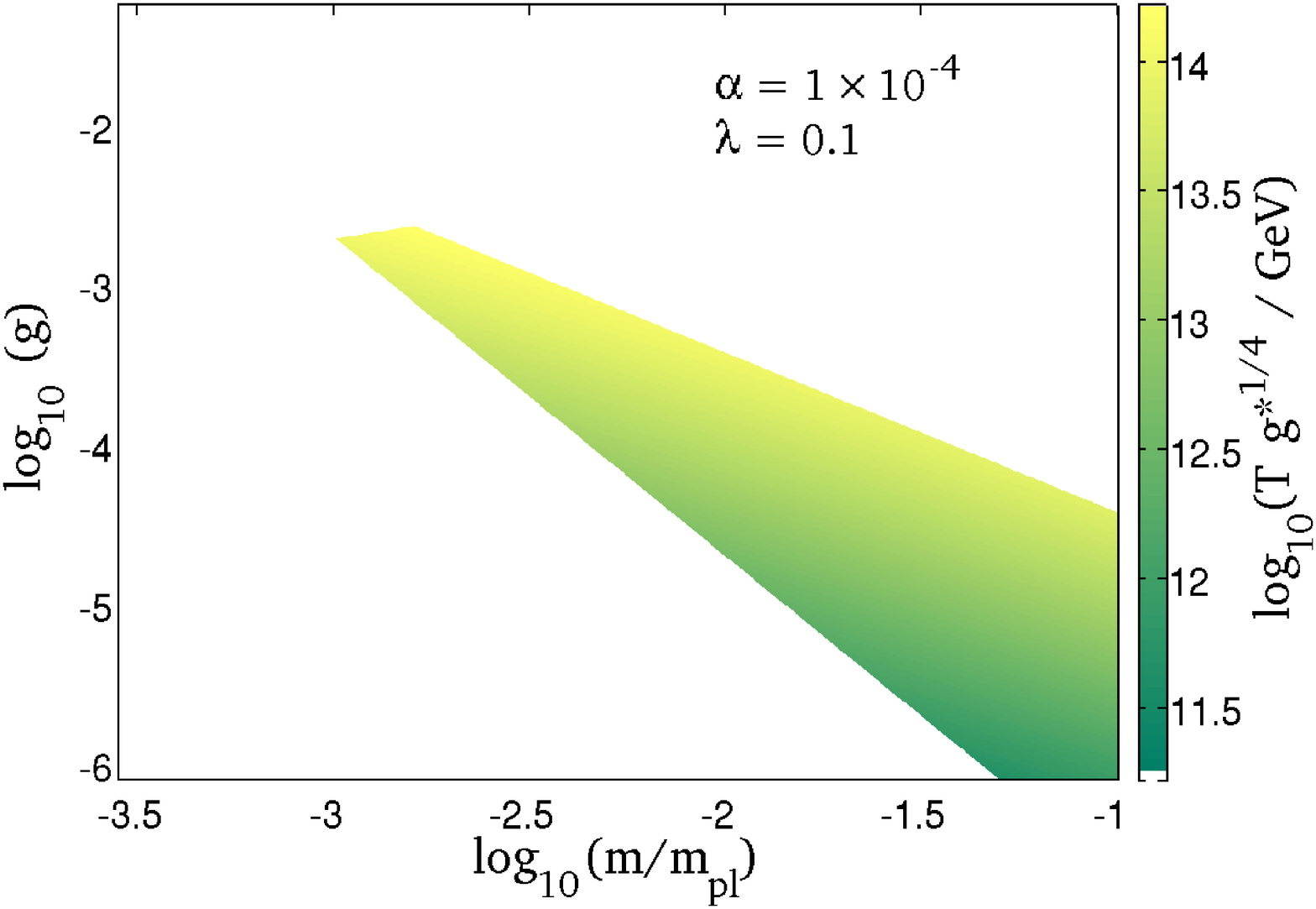}  
\end{tabular}
\caption{\label{figure3} The range for the reheating temperature
$T_{\rm RH}$ in the plane $g-m/m_{\rm pl}$, for $\lambda_{\chi} = 1$
(left-hand panel) and $\lambda_{\chi} = 0.1$ (right-hand
panel). Decreasing the value of $\lambda_{\chi}$ has the effect of
reducing the allowed region of parameter space. The shading density
indicates high (light) to low (dark) reheating temperature in the
allowed region of parameter space.}
\end{figure}

\end{widetext}
Treating the right-hand side as the average value over oscillations
of the $\chi$ field, we can rewrite it using $\rho_{\chi} = g^2 \phi^2
\langle \chi^2 \rangle$.  Writing these energy densities out explicitly,
including their evolution with scale-factor gives
\begin{equation}
\sqrt{{4 8\pi}}\lambda \left(\frac{a_x}{a_{{m}}} \right)^6
\left(\frac{a_{{m}}}{a} \right)^3 
\gg f(g,m) \left(\frac{m}{m_{\rm{pl}}} \right)^3 \frac{g^4}{4
  \lambda_\chi} \rm{e}^{-\Gamma_{\chi}(t-t_{{m}})} \,.
\end{equation}
The right-hand side will quickly become tiny when $t \sim
\Gamma^{-1}_{\chi}$. 
We need only consider the value of $a \propto t^{1/3}$ when this
occurs, and then $(a_{m}/a)^3 \simeq \Gamma_\chi t_m$. 
Using $\Gamma_\chi \simeq \alpha g m$ and
Eq. (\ref{fgm}) gives:  
\begin{equation}
\left(\frac{m_{\rm{pl}}}{m} \right)^{2/3}
\frac{g^{5/3}}{\lambda_{\chi}} \ll \sqrt{2} \alpha \lambda^{-1/6} \left(
\frac{a_{m}}{a_x} \right) \simeq 3\times 10^3 \alpha\,,  
\end{equation}
and thus unless the decay rate is tiny and the reheating period too
long, once we fulfill the other constraints this is practically always
fulfilled. In other words, having the parameter values such that the field
$\chi$ performs some oscillations around the minimum while the
expansion rate is dominated by the inflaton kinetic energy, those
oscillations will not backreact onto the motion of the inflaton
field.

\section{Range of Temperatures}

The constraints show there can be significant variation in the
resulting temperature. The energy density depends most sensitively on
$g$, and the range this parameter can take is severely constrained by
the other parameters. If $\lambda_{\chi}$ is too small, $g$ must be
made small enough to avoid interfering with the evolution of
$\phi$. If $m$ is made too small, $g$ must be made large enough to
ensure the field reacts while it is unstable. For instance, if
$\lambda_{\chi} \sim 1$ then $g$ can range from $g \sim 10^{-2}$ at its
largest (with $m/m_{\rm{pl}} \sim 10^{-3}$) to $g \sim 10^{-6}$ (with
$m/m_{\rm{pl}} \sim 10^{-1}$), assuming $m$ is kept below the Planck
scale. 
Decreasing $\lambda_{\chi}$ can constrain it further.

Reheating ends by the time $t \simeq \Gamma_\chi^{-1}$. Plugging
Eqs. (\ref{tre}), (\ref{fgm}), with $\Gamma_\chi \simeq \alpha gm$ and
$\lambda = 10^{-14}$, we have: 
\be
\rho_{\rm r} \simeq 8 \times 10^{-6} m_{\rm pl}^4 \alpha
\left(\frac{m}{m_{\rm pl}}\right)^{4/3} \frac{g^{11/3}}{\lambda_\chi} \ln
(\Gamma_\chi t_{m})^{-1}\,.
\label{rhorend}
\ee
Therefore, working for example with $\lambda_\chi \simeq 1$ and
$\alpha\simeq 10^{-4}$, we find the range 
\begin{equation}
g_*^{1/4}T_{\rm RH} \sim  ( 10^{11} - 10^{14} ) \ {\rm GeV} \,.
\end{equation}
One can also look at very weak coupling, where for
example $g \simeq 10^{-4}$, $\lambda_{\chi} \simeq 10^{-5}$, 
$\alpha \simeq 10^{-4}$, and
$m/m_{\rm pl} \simeq 10^{-2}$ gives $g_*^{1/4}T_{\rm RH} \simeq 3 \times
10^{13} {\rm GeV}$.
In Fig. {\ref{figure3}} we give the range for the reheating
temperature $T_{\rm RH}$ in
the plane $g-m/m_{\rm pl}$ for two values of $\lambda_\chi$. As we
decrease the value of the $\chi$ self-coupling, the allowed region in
the plane will be further reduced, and that would be the main effect 
on $T_{\rm RH}$.

\section{Transition to radiation domination}

The kinetic energy of the $\phi$ field is dropping quickly. Peebles
and Vilenkin noted that unless radiation domination occurred before the
kinetic energy reached the potential energy in $\phi$, the universe
would return to an inflationary regime from which it would never
recover. Our model has the same requirement. We will write $a_{\rm{end}}$ as
the scale-factor when reheating ends ($t \sim \Gamma_{\chi}^{-1}$) and the
radiation redshifts as an unsourced relativistic fluid.

The kinetic energy approaches the potential energy at a scalefactor
$a_*$, given by \cite{Peebles:1998qn} 
\begin{equation}
\frac{a_*}{a_{\rm{end}}} < \frac{a_x}{a_{\rm{end}}} \left(
\frac{m_{\rm{pl}}}{M} \right)^{4/3} \ln(m_{\rm{pl}}/M) \,.
\end{equation}
The energy in the $\phi$ field is
\begin{equation}
\rho_{\phi} \simeq\lambda m_{\rm{pl}}^4 (\Gamma_\chi t_m)^2 \left(
\frac{a_x}{a_{\rm{end}}}\right)^6 \left( \frac{a_{\rm{end}}}{a}\right)^6 \,,
\end{equation}
and the energy in the radiation is given in Eq. (\ref{rhorend}). 
The ratio is of order unity when the kinetic energy
reaches the energy in the radiation, at a scale-factor of $a_{\rm{r}}$. We 
require $a_{\rm{r}} < a_*$ to ensure radiation dominated expansion begins.
Then,
\begin{equation}
2 \times 10^{6} \frac{g^{7/3}}{\lambda_\chi}>  \left(\frac{M}{m_{\rm pl}} \right)^{8/3}\ln^{-2}\left(\frac{m_{\rm{pl}}}{M} \right)\,,
\end{equation}
which is easily satisfied for the values of the parameters we have
been considering, but does place a weak upper bound on
$\lambda_{\chi}$ for a given choice of $g$. 

Importantly, the energy in the potential of $\phi$ varies only very
slowly.  The exact scale-factor $a_{\rm{end}} < a_{\rm r} < a_*$, where
radiation transition occurs, matters very little in terms of its future
evolution: the variation of the field's potential energy in the range
in which this can happen is negligible, as the field is moving so slowly.
The same evolution outlined in \cite{Peebles:1998qn} therefore takes place,
with $\phi$ mimicking a cosmological constant and today's dark energy 
density obtained with $M \sim 10^6 \rm{GeV}$ .

\section{Domain Walls}

Typically in models with a symmetry breaking term, domain walls
inevitably form via the Kibble mechanism \cite{Kibble:1976sj}. Parts of
the universe causally separated have no way of being correlated.  When
the $\chi$ field becomes unstable, in some regions the field will move
to positive values, and in other regions to negative values. 
The domain walls created by the smooth
transition between these values are
generally a serious problem in many cosmological models, as they
can come to dominant the energy density of the 
universe \cite{vsbook}.

Fortunately, and unlike in many models with such an occurrence,
the symmetry is restored once $\phi > m$. Any walls formed will then
dissolve. However, this could leave some effect upon the amount of
reheating taking place within the regions of the domain wall,
potentially influencing large scale structure formation. As the wall
thickness is significantly smaller than a horizon, we do not expect
this to be a large effect, but warrants further investigation.

\section{Summary and future discussion}

By introducing an additional coupling to a second, self-interacting
scalar field, we can restore the traditional reheating mechanisms to
the scenario proposed by Peebles and Vilenkin \cite{Peebles:1998qn}. 
The symmetry breaking
term leaves the field temporarily unstable, which allows it to gain
significant amounts of energy. Using a phenomenological approach to
reheating, we have calculated the evolution of the relativistic
particles produced by the decay of this field.  We then described a
variety of constraints on the model parameters, ensuring that the
additional field does not interfere with the behaviour of the original
inflaton. The range in temperature produced is quite narrow if the new
field does not strongly self-interact, and the symmetry
breaking scale does not reach the Planck scale. But with a self
coupling of the order of $\lambda_\chi \gtrsim 0.01$, one has 
$ g_*^{1/4}T_{\rm RH} \sim 10^{11}- 10^{14}$ GeV. This allows us to suppress the
contribution of the gravitational waves at the time of BBN well below
the upper limit. 

With a model that tries to explain the evolution of the Universe from
early inflation to today dark energy domination, one of the issues 
to explain is the origin of the baryon asymmetry
\cite{Joyce:1996cp}. A possibility would be 
spontaneous baryogenesis through a derivative coupling of the inflaton
field to a matter current \cite{DeFelice:2002ir}. In our set-up, due
to the larger coupling of $\chi$ to the light degrees of freedom
(compared with the gravitational couplings), thermalization occurs
promptly after the start of reheating. This opens up the possibility
for example of having  leptogenesis during  or after reheating, by the
decay of the lightest right handed neutrino. The field $\chi$ being
quite heavy during reheating could
decay into the lightest right handed neutrino, or alternatively
the neutrinos could be thermally produced.      

The model could potentially form domain walls, often causing problems
in similar models. In this variation, we note they are transient and do
not interfere with future evolution of the universe. The possibility
exists their effect upon the reheating temperature will appear as
an imprint on large scale structure formation. 

The analyses of the reheating mechanism proposed in this letter has
been done at tree-level. But having coupled the quintessence field to
another scalar field, with a strength $g \gtrsim 10^{-6}$, the
question arises about the stability of the quintessential potential to
quantum corrections. Here we argue that indeed these corrections are
under control due to the decoupling theorem, and that the scenario is
not spoiled by quantum corrections, but leave the explicit calculation
for a future work.

Quantum corrections with the $\chi$ field running in the loops can give rise to
a potentially large correction to the quintessence potential, with the
one-loop correction given by \cite{Veff1}:
\be
\Delta V^{(1)}= \frac{1}{64 \pi^2} \sum_{i=\phi, \chi} m_i(\phi)^4
\left( \ln \frac{m_i^2(\phi)}{\mu^2} -\frac{1}{2} \right) \,,
\label{V1}
\ee where $\mu$ is the renormalization scale and $m_i(\phi)$ the field
dependent masses, with $m_\chi = g \phi$. So for the $\chi$ field we
have a very heavy state $m_\chi \sim g m_{\rm pl} \gg g M$ that is excited
in a universe with an energy density $\rho \ll \lambda M^4$. Given
that we do not 
have enough energy to excite such a heavy states, physically we can
expect that they decouple from the spectrum \cite{decoupling}, and
their contribution to the effective potential is highly
suppressed. However, the 1-loop effective potential as given in
Eq. (\ref{V1}) is computed using a mass independent renormalization
scheme that does not take into account threshold effects. The decoupling
would appear naturally when using instead a mass dependent
renormalization scheme \cite{mdrs}. To deal with this problem when
working with the effective potential one can use instead the
``improved'' effective potential \cite{improvedV} by replacing all
parameters (masses, couplings and field vevs) by their renormalized
values in both the tree level and one loop potential, and imposing the
physical condition that the potential does not depend on the
renormalization scale, $d V(\mu)/d \ln \mu |_{\mu=\mu^*}=0$. By
choosing the renormalization scale $\mu^*$ below any heavy mass
threshold in the model, heavy states decouple and the dependence on
$\mu$ is minimised. We are then left with the tree-level potential
again with all the parameters evaluated at $\mu^*$, \bea U
\left(\phi,\chi\right) &=& V(\phi, M(\mu^*), \lambda(\mu^*)) \nonumber
\\ && \!\!\!\!\!\!\!\!\!\!\!\!\!\!\!\!
+ \frac{g^2(\mu^*)}{2} \chi^2 \left(\phi^2 - m^2(\mu^*) \right) +
\frac{\lambda_\chi(\mu^*)}{4} \chi^4 \,.  \eea All possible quantum
corrections due to the heavy states are then encoded in the running of
the mass parameters and couplings through the renormalization group
equations, and are therefore expected to be under control.  In future
work, we plan to do a detailed analysis of quantum corrections and
decoupling in this model.

\vspace{-0.38cm}

\begin{acknowledgments}

\vspace{-0.32cm}

The work of M.B.G. is partially supported by
the M.E.C. under contract FIS2007-63364 and by the Junta de
Andaluc\'{\i}a group FQM 101.   
A.B. and B.M.J. acknowledge support from the STFC.

\end{acknowledgments}



\begin{thebibliography}{99}

\bibitem{Peebles:1998qn}
  P.~J.~E.~Peebles and A.~Vilenkin,
  Phys.\ Rev.\  D {\bf 59}, 063505 (1999).

\bibitem{chaotic}
  A.~D.~Linde,
  Phys.\ Lett.\  B {\bf 129} (1983) 177.

\bibitem{Spokoiny:1993kt}
  B.~Spokoiny,
  Phys.\ Lett.\  B {\bf 315} (1993) 40.

\bibitem{Ford:1986sy}
  L.~H.~Ford,
  Phys.\ Rev.\  D {\bf 35} (1987) 2955.

\bibitem{Albrecht:1982mp}
  A.~J.~Albrecht, P.~J.~Steinhardt, M.~S.~Turner and F.~Wilczek,
  Phys.\ Rev.\ Lett.\  {\bf 48}, 1437 (1982);
A. D. Dolgov and A. D. Linde, Phys. Lett. B{\bf 116} (1982) 329;
  L.~F.~Abbott, E.~Farhi and M.~B.~Wise,
  Phys.\ Lett.\  B {\bf 117} (1982) 29.


\bibitem{Starobinsky:1979ty}
  A.~A.~Starobinsky,
  JETP Lett.\  {\bf 30} (1979) 682
  [Pisma Zh.\ Eksp.\ Teor.\ Fiz.\  {\bf 30} (1979) 719].

\bibitem{Allen:1987bk}
  B.~Allen,
  Phys.\ Rev.\  D {\bf 37} (1988) 2078.

\bibitem{Sahni:1990tx}
  V.~Sahni,
  Phys.\ Rev.\  D {\bf 42} (1990) 453.

\bibitem{Sahni:2001qp}
  V.~Sahni, M.~Sami and T.~Souradeep,
  Phys.\ Rev.\  D {\bf 65} (2002) 023518.

\bibitem{Giovannini:1998bp}
  M.~Giovannini,
  Phys.\ Rev.\  D {\bf 58} (1998) 083504.


\bibitem{braneworld}
  R.~Maartens, D.~Wands, B.~A.~Bassett and I.~Heard,
  Phys.\ Rev.\  D {\bf 62} (2000) 041301.
  E.~J.~Copeland, A.~R.~Liddle and J.~E.~Lidsey,
  Phys.\ Rev.\  D {\bf 64} (2001) 023509.
  G.~Huey and J.~E.~Lidsey,
  Phys.\ Lett.\  B {\bf 514} (2001) 217.


\bibitem{instantpre}
  A.~H.~Campos, H.~C.~Reis and R.~Rosenfeld,
  Phys.\ Lett.\  B {\bf 575}, 151 (2003).
  M.~Sami and V.~Sahni,
  Phys.\ Rev.\  D {\bf 70} (2004) 083513.

\bibitem{curvreh}
  B.~Feng and M.~z.~Li,
  Phys.\ Lett.\  B {\bf 564} (2003) 169.
  K.~Dimopoulos,
  Phys.\ Rev.\  D {\bf 68} (2003) 123506.
  A.~R.~Liddle and L.~A.~Urena-Lopez,
  Phys.\ Rev.\  D {\bf 68} (2003) 043517.

\bibitem{BIreh}
  M.~Sami, N.~Dadhich and T.~Shiromizu,
  Phys.\ Lett.\  B {\bf 568} (2003) 118.

\bibitem{abwhepp}
A. Berera, Pramana {\bf 72} (2009) 169.

\bibitem{Linde:1993cn}
  A.~D.~Linde,
  Phys.\ Rev.\  D {\bf 49}, 748 (1994).



\bibitem{preheating}
  L.~Kofman, A.~D.~Linde and A.~A.~Starobinsky,
  Phys.\ Rev.\ Lett.\  {\bf 73}, 3195 (1994).
  L.~Kofman, A.~D.~Linde and A.~A.~Starobinsky,
  Phys.\ Rev.\  D {\bf 56}, 3258 (1997).
 

\bibitem{Kibble:1976sj}
  T.~W.~B.~Kibble,
  J.\ Phys.\ A  {\bf 9}, 1387 (1976).

\bibitem{vsbook}
A. Vilenkin and E. P. Shellard,
``Cosmic Strings and Other Topological Defects'',
Cambridge University Press (1994).

\bibitem{Joyce:1996cp}
  M.~Joyce,
  Phys.\ Rev.\  D {\bf 55} (1997) 1875.
\bibitem{DeFelice:2002ir}
  A.~De Felice, S.~Nasri and M.~Trodden,
  Phys.\ Rev.\  D {\bf 67}, 043509 (2003).

\bibitem{Veff1}
S. Coleman and E. Weinberg, Phys.\ Rev.\  D {\bf 7}, 1888 (1973);
S. Weinberg, Phys.\ Rev.\  D {\bf 7}, 2887 (1973).    

\bibitem{decoupling}
K. Symmanzik, Comm. Math. Phys. {\bf 34} 7 (1973);
T. Appelquist and J. Carazzone,   Phys.\ Rev.\  D {\bf 11}, 2856 (1975). 

\bibitem{mdrs}
 H.~Georgi and H.~D.~Politzer,
  Phys.\ Rev.\  D {\bf 14} (1976) 1829.

\bibitem{improvedV}
M.~Bando, T.~Kugo, N.~Maekawa and H.~Nakano,
  Phys.\ Lett.\  B {\bf 301} (1993) 83;
M.~Bando, T.~Kugo, N.~Maekawa and H.~Nakano,
  Prog.\ Theor.\ Phys.\  {\bf 90} (1993) 405;
  Nucl.\ Phys.\  B {\bf 553} (1999) 511. 




\end{thebibliography}
\end{document}